\documentstyle [12pt]{article}
\begin{document}
\title{The Boltzmann/Shannon entropy \\
       as a measure of correlation}

\author{John H. Van Drie\\
        www.johnvandrie.com \\
        Kalamazoo, MI 49008  USA}

\maketitle

\begin{abstract}
It is demonstrated that the entropy of statistical mechanics and of information theory,
$S({\bf p}) = -\sum p_i \log p_i $ may be viewed as a measure of correlation.  
Given a probability distribution
on two discrete variables, $p_{ij}$, we define the correlation-destroying transformation
$C: p_{ij} \rightarrow \pi_{ij}$, which creates a new distribution on those same variables in
which no correlation exists between the variables, i.e. $\pi_{ij} = P_i Q_j$.  It is then shown
that the entropy obeys the relation $S({\bf p}) \leq S({\bf \pi}) = S({\bf P}) + S({\bf Q})$, i.e. 
the entropy is non-decreasing under these correlation-destroying transformations. 
\end{abstract}

The concept of correlation has underlain statistical mechanics from its inception.  Maxwell
derived his velocity distribution law \footnote{Maxwell, J.C., {\it Phil. Soc.}, 1860} by asserting
that such a distribution $\Phi(\vec{v})$ for an ideal gas should obey two 
properties:  (1) the velocity distribution along each axis should be uncorrelated, i.e.
$$
\Phi(\vec{v}) d^3\vec{v} = (\phi(v_x) dv_x) (\phi(v_y) dv_y) (\phi(v_z) dv_z)
$$
and (2) the velocity distribution should
show no preferred orientation, $\Phi(\vec{v}) d^3\vec{v} = f(v) d^3\vec{v}$, 
where v is the norm of $\vec{v}$.  He
showed these two assumptions led to the velocity distribution 
$\Phi(\vec{v}) d^3\vec{v} = \exp (-\alpha v^2) d^3\vec{v}$,
where $\alpha$ is a positive constant (later shown by Boltzmann to be equal to $\frac{m}{2kT}$). 
(This reduces to the more familiar form by writing this expression in polar coordinates,
$\Phi(v_r,v_{\theta},v_{\phi}) d^3\vec{v}= \exp (-\alpha v_r^2) v_r^2 dv_r dv_{\theta} dv_{\phi}$).

Inspired by the observations of Ochs \footnote{Ochs, W., {\it Rep. Math. Phys.}, {\bf 9}, 135 (1976)}, 
we would like to show that
the Boltzmann/Shannon formula for entropy may be viewed as a measure of correlation, by showing that
for a class of transformations which destroys correlations between variables in a probability
distribution, the Boltzmann/Shannon formula for the entropy is non-decreasing.

Suppose that we have a set of states, $\{ X_{ij} \}$, indexed by their values along two distinct, 
discrete variables A and B, where i runs over the set of n discrete states of A, and j runs over the
m set of discrete states of B.  Consider a probability distribution over these states, $p_{ij}$.  If 
A and B are uncorrelated, there exist some $P_i$ and $Q_j$ such that $p_{ij} = P_i Q_j \forall i,j$.
In this case, it is apparent that the entropy obeys the property S(p) = S(P) + S(Q).

However, in general, this will not be true.  But, given an arbitrary $p_{ij}$, we can define the following
transformation, which in effect destroys the correlations between its dependence on A and on B:
\begin{eqnarray}
C: & p_{ij} \rightarrow \pi_{ij} \\
\pi_{ij} & = & P_i Q_j \\
P_i & = & \sum_{j=1}^m p_{ij} \\
Q_j & = & \sum_{i=1}^n p_{ij}
\end{eqnarray}

We assert that the entropy is non-decreasing under such transformations, i.e. 
\begin{equation}
S(p) \leq S(\pi)    \label{eq:biggie}
\end{equation}

To demonstrate this assertion, we need first to demonstrate a fundamental property of the Boltzmann/Shannon 
entropy formula, {\it the averaging property}.  Given a set of states $\{Y \}_{k=1}^N$, and two probability
distributions defined over these states, $G = \{g_k\}_{k=1}^N$ and $H = \{h_k\}_{k=1}^N$, one may construct
a third distribution, $U = \{u_k\}, u_k = \frac{1}{\alpha + \beta}(\alpha g_k + \beta h_k)$, the weighted
average of G and H, where $\alpha$ and $\beta$ are arbitrary real values.  We assert that
\begin{equation}
S(U) \geq \frac{1}{\alpha + \beta} ( \alpha S(G) + \beta S(H) )
\end{equation}
This assertion can be demonstrated by observing that the similar inequality holds term-by-term in the sum.
Defining $\sigma(x) = - x \ln x$, the averaging property will hold if
\begin{equation}
\sigma(u_k) \geq \frac{1}{\alpha + \beta} ( \alpha \sigma(g_k) + \beta \sigma(h_k) ) 
\end{equation}
This property of $\sigma(x)$ follows from it being concave everywhere over the domain of interest, $x \in [0,1]$,
i.e. $\sigma''(x) < 0$.

Note, too, that a consequence of the averaging property of entropy is that, given a set of different distributions over
$\{Y \}_{k=1}^K$, $Z^1, Z^2, Z^3, \dots Z^K$, and a set of weights $w_i, \sum_i w_i = 1$, 
that the entropy of $\bar{Z} = \sum_k w_k Z^k$, 
the weighted average of all these distributions, obeys the following property:
\begin{equation}
S(\bar{Z}) \geq \sum_k w_k S(Z^k) 
\end{equation}
Returning to the fundamental assertion, equation ~\ref{eq:biggie}, 
this may be demonstrated by recalling a fundamental
property of the Boltzmann/Shannon entropy formula, one which Shannon \footnote{C. Shannon and W. Weaver,
{\it The Mathematical Theory of Communication},Urbana: Univ. of Ill. Press, 1949} 
took not as a derived property but rather an
axiomatic property that an entropy functional must have:  If we decompose the distribution
$p_{ij}$ into a two stages, where initially we distribute among the states over A by the distribution
$P_i$, and next we distribute among the states of B by the distribution $\zeta^{(i)}_j$, such that
$p_{ij} = P_i \zeta^{(i)}_j$, the entropy obeys the formula
\begin{equation}
S(p) = S(P) + \sum_i P_i S(\zeta^{(i)})
\end{equation}
In the general case, each of the distributions $\zeta^{(i)}$ will be different, i.e. the variables
A and B are correlated.  The distribution $Q_j$ above represents a weighted average of the $\zeta^{(i)}$'s,
weighted by $P_i$, i.e.
\begin{equation}
Q_j = \sum_k P_k \zeta^{(k)}_j
\end{equation}
Hence, the Shannon axiom with the averaging property of the entropy leads to the desired assertion:
\begin{eqnarray}
S(p)  & = &  S(P) + \sum_k P_k S(\zeta^{(k)})  \\
S(p) & \leq & S(P) + S(Q) \\
S(p) & \leq &  S(\pi)   
\end{eqnarray}

Jaynes \footnote{Jaynes, {\it Phys. Rev.}, {\bf 106}, 620 (1957)} showed how Shannon's theory could be merged
with statistical mechanics, leading to the conceptualization of the thermodynamic principle of
maximum entropy as a principle expressing that the distribution of energy among the microstates
should be that distribution which is least-biased, given the constraint of a specified temperature.

Viewing entropy $-k \sum_k p_i \ln p_i$ as a measure of (lack of) correlation provides a new twist to
Jaynes' perspective.  One may say that the equilibrium distribution is that distribution which
is least-correlated given the constraint(s).  The Second Law of Thermodynamics may be rephrased to
state that correlations are highly unlikely to arise spontaneously, and that the natural course of
evolution of a system is one in which correlations diminish.  

Thinking about entropy as a measure of correlation leads to a key implicit assumption in both
Boltzmann's theory and Shannon's theory:  the individual (micro)states are assumed to be
uncorrelated.  This hails back to Laplace's balls-in-urns, where the probability of finding
a ball in a given urn is generally uncorrelated with the probability of finding a ball in a different
urn.  If the probability of occupation of the states are intrinsically correlated, the maximum
entropy distribution cannot be viewed as the correct, least-biased distribution.  In communication
theory and statistical mechanics, this assumption may in certain circumstances be valid, 
but where this assumption breaks down severely 
is the case when we attempt to take the limit to a continuous set of states.
If one constructs a discrete set of bins from an intrinsically continuous variable, the degree
of bin-bin correlation grows as these bins become steadily finer.  This leads to the question
'what is the correct measure of entropy for a continuous distribution?'.

Dill \footnote {Dill, K.A.,{\it J. Biol.Chem.}, {\bf 272}, 701 (1997)} points out that these
issues of correlation and additivity pervade our thinking about the fundamental aspects
of chemical and biological phenomena.  He highlights some of the pitfalls which one may encounter
in settings for which an assumption of non-correlation may not be valid.

\end{document}